\documentclass[useAMS,usenatbib,usegraphicx]{mn2e}

\title[Evolution of Multi-mass Star Clusters]{Effects of External
Tidal Field on the Evolution of Multi-mass Star Clusters}
\author[K. H. Lee et al.]{K. H. Lee,$^{1,2}$\thanks{E-mail:
khlee@arcsec.sejong.ac.kr} H. M. Lee$^{3}$ and H. Sung$^{1}$\\
$^{1}$Astrophysical Research Center for the Structure and
Evolution of the Cosmos, Sejong University, Seoul 143-747, Korea\\
$^{2}$Centre for Astrophysics \& Planetary Science, School of
Physical Sciences, University of Kent, Canterbury, Kent CT2 7NR,
UK\\
$^{3}$Astronomy Program, SEES, Seoul National University, Seoul
151-742, Korea}

\begin{document}


\pagerange{\pageref{firstpage}--\pageref{lastpage}} \pubyear{2002}

\maketitle

\label{firstpage}

\begin{abstract}
We present N-body simulations of realistic globular clusters
containing initial mass function in the galaxy to study effects of
tidal field systematically on the properties of outer parts of
globular clusters. Using NBODY6 which takes into account the
two-body relaxation correctly, we investigate general evolution of
globular clusters in galactic tidal field. For simplicity, we have
employed only spherical components (bulge and halo) of the galaxy.
Total number of stars in our simulations was about 20,000. All
simulations were done for several orbital periods in order to
understand the development of the tidal tails. In our scaled down
models, the relaxation time is sufficiently short to show the mass
segregation effect, but we did not go far enough to see the
core-collapse, and the fraction of stars lost from the cluster at
the end of simulation is only about $\sim 10 \%$. The radial
distribution of extra-tidal stars can be described by a power law
with a slope around $-3$. The direction of tidal tails are
determined by the orbits and locations of the clusters. We find
that the length of tidal tails increases towards the apogalacticon
and decreases towards the perigalacticon. This is an
anti-correlation with the strength of the tidal field, caused by
the fact that the the time-scale for the stars to respond to the
potential is similar to the orbital time-scale of the cluster.
When the length of tidal tails decreases some of the stars in the
tidal tails are recaptured by the host cluster. From the
investigation of velocity anisotropy of the model clusters, we
find that in the early stages of globular cluster evolution the
clusters have radial anisotropy in the outermost parts, while
clusters are nearly isotropic in the cental region. The radial
anisotropy decreases with time.
\end{abstract}


\section{Introduction}

Dynamical evolution of globular clusters is induced by a variety
of internal and external processes. The main processes dominating
the evolution of clusters are two-body relaxation and tidal
shocks. Two-body relaxation makes the velocity distribution of
stars towards Maxwellian and high-velocity stars gain enough
energy to escape from the cluster (Spitzer \& Thunan 1972). The
external tidal shocks also accelerate the destruction of clusters
(see, for example, Gnedin, Lee \& Ostriker 1999). Clusters
experience tidal shocks when they pass through the Galactic disk
or close to the Galactic bulge. The escaped stars may remain in
the vicinity of the cluster and form tidal tails for a while. Many
observational studies show that the signs of the existence of
tidal tails around globular clusters (Grillmair et al. 1995;
Grillmair et al. 1996; Holland, Fahlman \& Richer 1997; Lehman \&
Scholz 1997; Leon, Bergond \& Vallenari 1999; Leon, Meylan \&
Combes 2000; Testa et al. 2000; Odenkirchen et al. 2001, 2003;
Sohn et al. 2003; Lee et al. 2003, 2004). The features of tidal
tails are thought to be related to the orbit and location of the
cluster and the Galactic potential. However the links between the
properties of tidal tails and the cluster dynamics have not been
fully understood yet. Combes et al. (1999) presented extensive
N-body simulations of globular clusters, and showed the tidal
effects quantitatively and geometrically. The main finding of
Combes et al. (1999) is that there exist two giant tidal tails
around the globular cluster along cluster's orbit. They also
showed that the density distribution just outside the tidal
boundary follows power laws to radius, and the extra-tidal stars
form clumps which are the tracers of the strongest gravitational
shocks. However, they used the PM method for the calculation of
the gravitational potential: this method cannot take into account
the two-body relaxation effect on the internal evolution of the
clusters. Yim \& Lee (2002) examined the general evolution of the
globular clusters using NBODY6 which properly takes into account
the effects of two-body relaxation. They showed that the density
profiles of the clusters appear to become somewhat shallower just
outside the tidal boundary as observed in many clusters and the
directions of the tidal tails depend on the location in the galaxy
as well as the cluster orbit.

Dynamical evolution of many body systems can be studied by various
methods. The most desirable method to study dynamical evolution of
a globular cluster is a direct N-body integration. However, this
method becomes almost impossible to use as $N$ becomes large,
where $N$ is the number of stars, since the computational time
increases with $N^3$. The required value of $N$ (of order $10^6$)
is still beyond the capability of currently available computers.
Fokker-Planck method has been commonly used as an alternative to
direct N-body integration method. The main advantage of
Fokker-Planck method is that it does not require much computing
power. However, in this method, the shape of the cluster is
restricted to sphere (or oblate spheroid for rotating models, see
Einsel \& Spurzem 1999) and the external gravitational field can
be accounted for only approximately. To investigate tidal tails of
globular clusters in the external potential realistically, we use
N-body method in this study, but with significantly smaller number
of $N$ than realistic globular clusters. We will focus on the
behavior of stars especially in the outer parts of the globular
clusters. This paper is organized as follows. In Sect. 2, we
briefly describe the method and models. And the results of
simulations are described in Sect. 3. The final section summarizes
out major findings.

\section[]{Method and Model}

\subsection{NBODY6++}

N-body simulations for this study have been carried out using
NBODY6++ which is a parallelized version of NBODY6 developed by
Aarseth to use parallel super computer (Spurzem 1999). We modified
the program to include external potential, which is described in
\S 2.2. The code uses a direct summation method in computing the
gravitational potential and adopts Hermite integration scheme
which is simple to use and highly effective for advancing the
single particles. The main idea of Hermite integration scheme is
to employ a fourth-order force polynomial but now the two first
terms are evaluated by explicit summation over all $N$ particles,
thereby enabling two corrector terms to be formed. The code also
adopts an Ahmad-Cohen neighbor scheme (Ahmad \& Cohen 1973) which
is based on the idea of separating the total force on a particle
into two components,
\[
F = F_{irr} + F_{reg}.
\]
where $F_{irr}$ is the irregular force caused by relatively nearby
stars and $F_{reg}$ is the regular force from rather distant
stars. The irregular force varies with time rather rapidly while
the regular force changes slowly. Since the external force due to
the host galaxy is expected to vary rather slowly, we regarded
such force as a regular one. Each part is represented by the
high-order force polynomial on its own time-step. The neighbor
force is updated on a time-scale determined by the local
fluctuations, whereas the smoother regular component due to all
other members need only be recalculated on a longer time-scale,
with intermediate values obtained by prediction. At each regular
time-step a new neighbor list is determined using a given neighbor
radius for each particle. It also includes the formation of
binaries via three-body processes and their interactions with
other stars. The main force loop in NBODY6 is decomposed in three
independent DO-loops, advancing all the regularized binaries (1),
computing new irregular forces and applying the corrector due to
them (2), and computing new regular forces, new neighbor lists and
applying the regular force corrections (3). The entire content of
the routines relevant to (2) and (3) have been parallelized in
NBODY6++. This is all done fully in parallel for the different
particles due in this block (Spurzem \& Baumgardt 2002).  Detailed
mechanisms of NBODY6 are described in Aarseth (1999, 2004) and
references therein.

\subsection{The Model Galaxy}

The gravitational potential of the Galaxy is generated by stars,
gas, and dark matter. There is no single numerical method that can
describe the evolution of the Galaxy since there are many
components that react differently. Further, we expect that the
time-scale for the variation of Galactic potential would be much
longer than the orbital time-scale of the clusters. Therefore the
Galactic potential is assumed to be independent of time for
simplicity. The main components of the Galaxy are central bulge,
disk, halo, and bar. In the present study, the model Galaxy is
assumed to be composed of two components: halo and bulge, to focus
on the effects from spherical components only. We have used simple
model potentials following Lee et al. (1999). The halo component,
which gives rise to the flat rotation curve at large radii, is
assumed to have a logarithmic potential,
\[
\Phi(r)_{halo} = {1 \over 2} v_0^2 \ln (R_c^2 + r^2) + const,
\]
where $R_c$ is the halo core radius and $v_0$ is the constant
rotation velocity at large $r$. As for the bulge component, we
have assumed the Plummer model to allow a steep outward velocity
increase from the Galactic centre (Binney \& Tremaine 1987),
\[
\Phi(r)_{bulge} = - {GM_{bulge} \over \sqrt{r^2 + r_c^2}},
\]
where $r_c$ is a parameter that controls the size of bulge. The
bulges of real galaxies can be better represented by de
Vaucouleurs's $R^{1/4}$ law, but we have chosen the Plummer model
for simplicity. Our model have $r_c \simeq 230 \sim 240$ pc. In
our numerical calculations all physical quantities are given with
dimensionless units. The unit of length is chosen to be $R_{sc} =
10$ kpc. The total mass of the Galaxy within 10 kpc is assumed to
be $M_{10} = 1.24 \times 10^{11}M_\odot$ (Caldwell \& Ostriker
1981). The unit of velocity is $v_{sc} = (GM_{10}/R)^{1/2} \simeq
230.7$km/s. In these units, the rotation velocity $v_0$ at the
distance of 10 kpc is 0.92 and $M_{bulge} = 0.08$, $R_c = 0.7$.

\begin{figure}
\includegraphics [width=84mm] {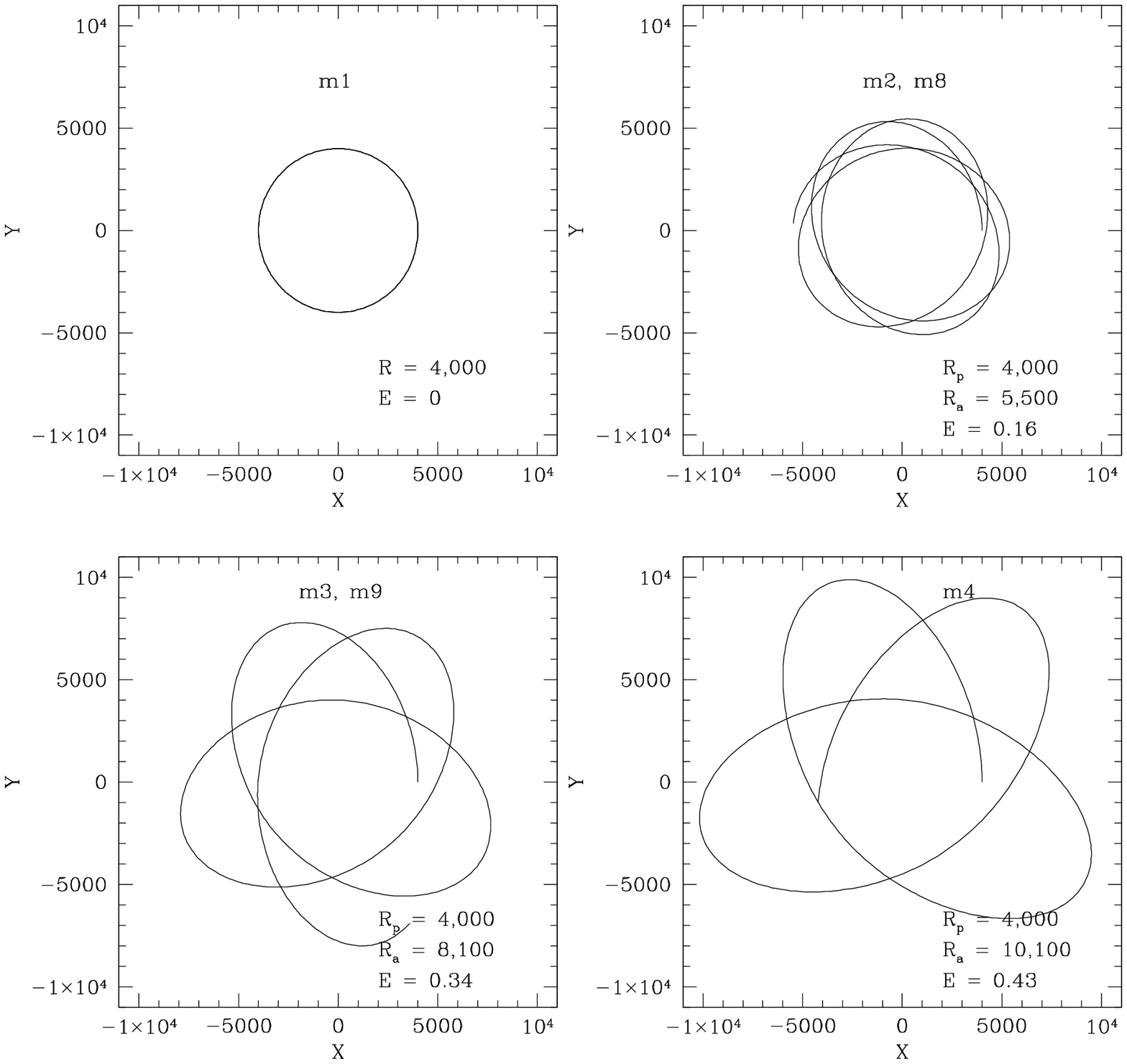}
\includegraphics [width=84mm] {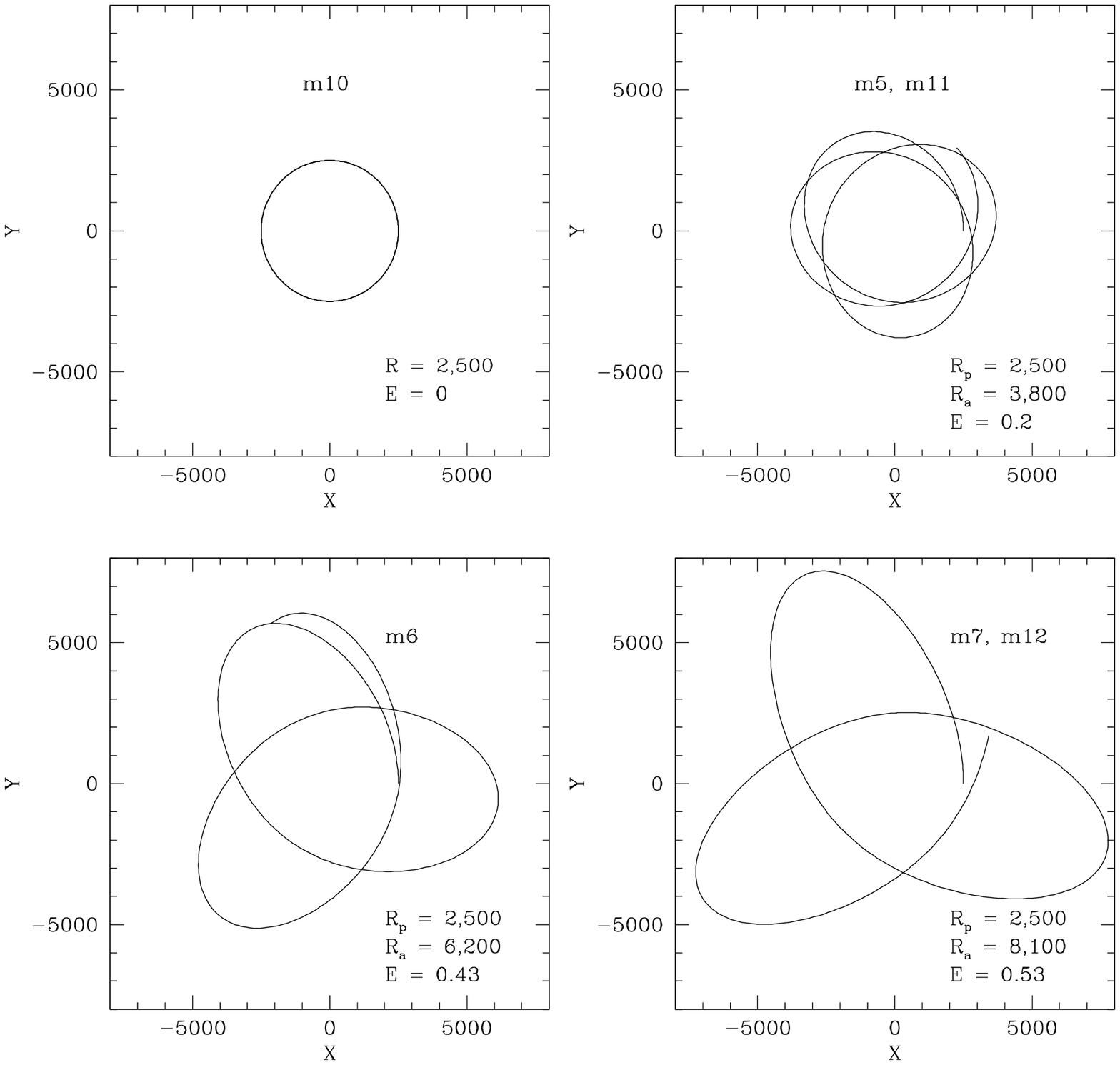}
 \caption{Orbits of model clusters. $R_p$ is
distance of the perigalacticon, $R_a$ is distance of the
apogalacticon, and E is the ellipticity of orbits.}
\end{figure}

\subsection{Initial Conditions of Globular Clusters and Their Orbits}

We have used the King model with W$_0$ = 5 and W$_0$ = 7 as
initial models, where W$_0$ is the scaled central potential, which
determines the degree of central concentration (King 1966). The
density profiles of all mass components are assumed to be the same
at the beginning (i.e. no initial mass segregation). We choose the
initial mass spectrum to be a simple power law
\[
dN(m) \propto m^{-\alpha}dm,
\]
where $dN(m)$ is the number of stars with masses between $m$ and
$m+dm$, and we have chosen $\alpha = 2.35$ which represents a
Salpeter initial mass function. The range of mass in our models is
from 1 to 10 $\rm M_\odot$. The total mass of a cluster is $M =
4.2\times 10^4 M_\odot$, and the number of stars is $N = 18,809$.
We have ignored the effects of stellar evolution in order to
concentrate on the pure dynamical processes and also for
simplicity. We have carried out a total of 12 simulations with
different cluster orbits. For models with elongated orbits, we
defined the ellipticity as,
\[
E = {{R_a - R_p} \over {R_a + R_p}}
\]
where $R_a$ and $R_p$ are apogalacticon and perigalacticon,
respectively. The parameters of models are listed in Table~1, and
the cluster orbits are shown in Fig.~1. We have removed the stars
from our simulations when the distance from the cluster centre
exceeds five times of the tidal radius. The tidal radius $r_t$ of
the cluster is defined as,
\[
r_t \equiv \left( {M_c\over 2M_G}\right)^{1/3} R_a,
\]
where $M_c$ is the cluster mass, and $M_G$ is the galactic mass
within radius $R_a$. Here we have assumed that the tidal radius is
determined by the tidal force at the apogalacticon. We are mostly
interested in the development of tidal tails, and thus we do not
follow the cluster until its destruction.

\begin{table}
\centering
 \begin{minipage}{140mm}
  \caption{Parameters of the cluster models and orbits.}
  \begin{tabular}{@{}llrrrrlrlr@{}}
  \hline
Model & W$_0$ & $r_c$ & $r_t$ & $r_{peri}$ &
$r_{apo}$ & E & Period \\
 & & (pc) & (pc) & (kpc)& (kpc) & & (Myr) \\
  \hline
m1 & 5 & 2.5 & 25 & 4 & 4 & 0 & 167 \\
m2 & 5 & 2.5 & 25 & 4 & 5.5 & 0.16 & 242 \\
m3 & 5 & 2.5 & 25 & 4 & 8.1 & 0.34 & 294 \\
m4 & 5 & 2.5 & 25 & 4 & 10.1 & 0.43 & 331 \\
m5 & 5 & 1.5 & 16 & 2.5 & 3.8 & 0.20 & 191 \\
m6 & 5 & 1.5 & 16 & 2.5 & 6.2 & 0.43 & 238 \\
m7 & 5 & 1.5 & 16 & 2.5 & 8.1 & 0.53 & 270 \\
m8 & 7 & 0.8 & 25 & 4 & 5.5 & 0.16 & 242 \\
m9 & 7 & 0.8 & 25 & 4 & 8.1 & 0.34 & 294 \\
m10 & 7 & 0.6 & 16 & 2.5 & 2.5 & 0 & 105 \\
m11 & 7 & 0.6 & 16 & 2.5 & 3.8 & 0.20 & 191 \\
m12 & 7 & 0.6 & 16 & 2.5 & 8.1 & 0.53 & 270 \\
\hline
\end{tabular}
\end{minipage}
\end{table}

\subsection{Time-Scale Consideration}

The number of stars of around $N\sim 20,000$ used in the present
study is clearly much smaller than that of realistic globular
clusters. Different dynamical processes operate on different
time-scales. The cluster's orbit depends only on the orbital
parameters (such as the ellipticity and peri- or apo-galacticon
distance), and not on $N$. The orbital time of the stars near the
tidal boundary is also very close to the orbital period of the
cluster. However, the dynamical evolution such as core-collapse
and mass segregation take place in two-body relaxation time-scale.
Since the relaxation time depends on local variables such as
stellar density and velocity dispersion, a single representative
value was defined as `half-mass relaxation time' by Spitzer \&
Hart (1975) as
\[
t_{rh} \equiv 0.138 {N^{1/2} r_h^{3/2} \over m^{1/2} G^{1/2} \ln
(0.4N)},
\]
where $r_h$ is the radius within which the enclosed mass is the
half of the total mass and $m$ is the mass of the individual
stars. For clusters with mass function, $m$ is not a well defined
parameter, but we can simply replace it by the mean mass although
there is no clear reason (see Lee \& Goodman 1995 for more
discussion). As we can see from the above equation, $t_{rh}
\propto N^{1/2}$. This means that the dynamical evolution of our
models takes place much faster than real globular clusters for a
given orbital period. The initial half-mass relaxation time of our
model is about $3.7 \times 10^8$ yr which is much shorter than
those of real globular clusters. The evaporation time-scale is
also proportional to the half-mass relaxation time (Henon 1961,
Lee \& Ostriker 1986), and thus we expect the development of the
tidal tails would be also relatively fast. However, since our
simulation would last for only up to about $10^9$ years, the
entire evolution we observe from the simulation could be
comparable to that of real globular clusters.

The amount of time for stars remaining in the tidal tails is
expected to be of order of orbital time-scale of the cluster
(e.g., Lee \& Ostriker 1986), which is not related to the number
of stars. Therefore, the dynamical features in the tidal tails of
our simulation would not be affected by the fact that we used the
smaller number of stars than realistic clusters. In Table 1, we
have listed the radial period which is the time from
perigalacticon to next perigalacticon of orbits in units of years
for our models.

\begin{figure}
\includegraphics  [width=84mm] {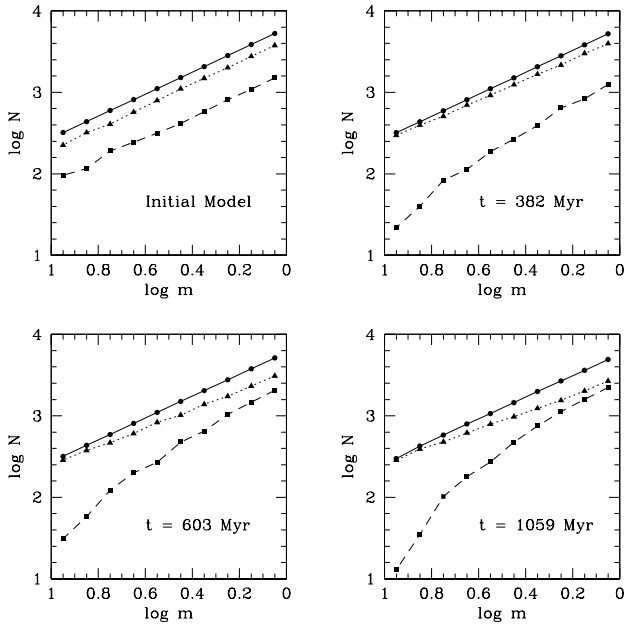}
 \caption{Mass functions of model m3 at a few selected epochs.
Filled circles, MF from the whole stars; filled triangles, inner
region; filled squares, outer region. We can see that mass
segregation effect increases with time.}
\end{figure}

\begin{figure}
\includegraphics [width=84mm] {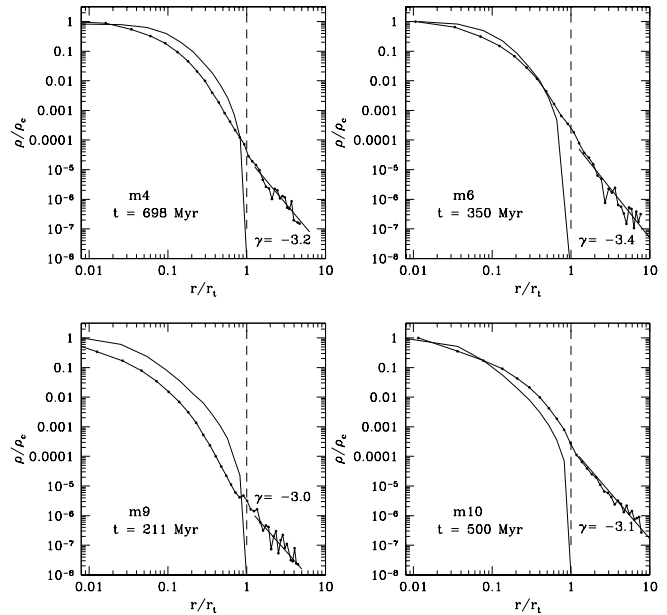}
 \caption{Radial density profile of models m4, m6, m9, and m10.
The initial profiles are shown for comparison. Note the clear
indication of tidal tails. The average power law gradient is
$-3.2$.}
\end{figure}

\begin{figure}
\includegraphics [width=84mm] {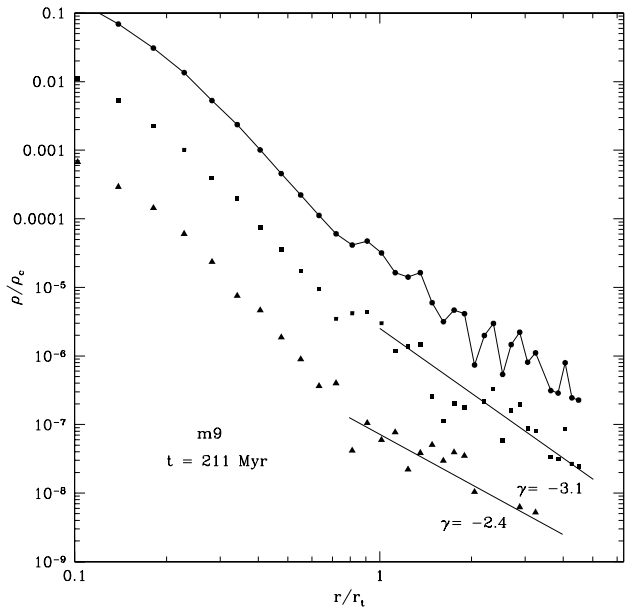}
 \caption{Fit of power laws to the external profile of model m9.
Triangle, the profile of massive stars; filled squares, the
profile of less massive stars; filled circles, the profile of all
stars; solid lines are the fitting power laws. The profile of less
massive stars shows a steeper slope. }
\end{figure}

\begin{figure*}
\includegraphics [width=180mm] {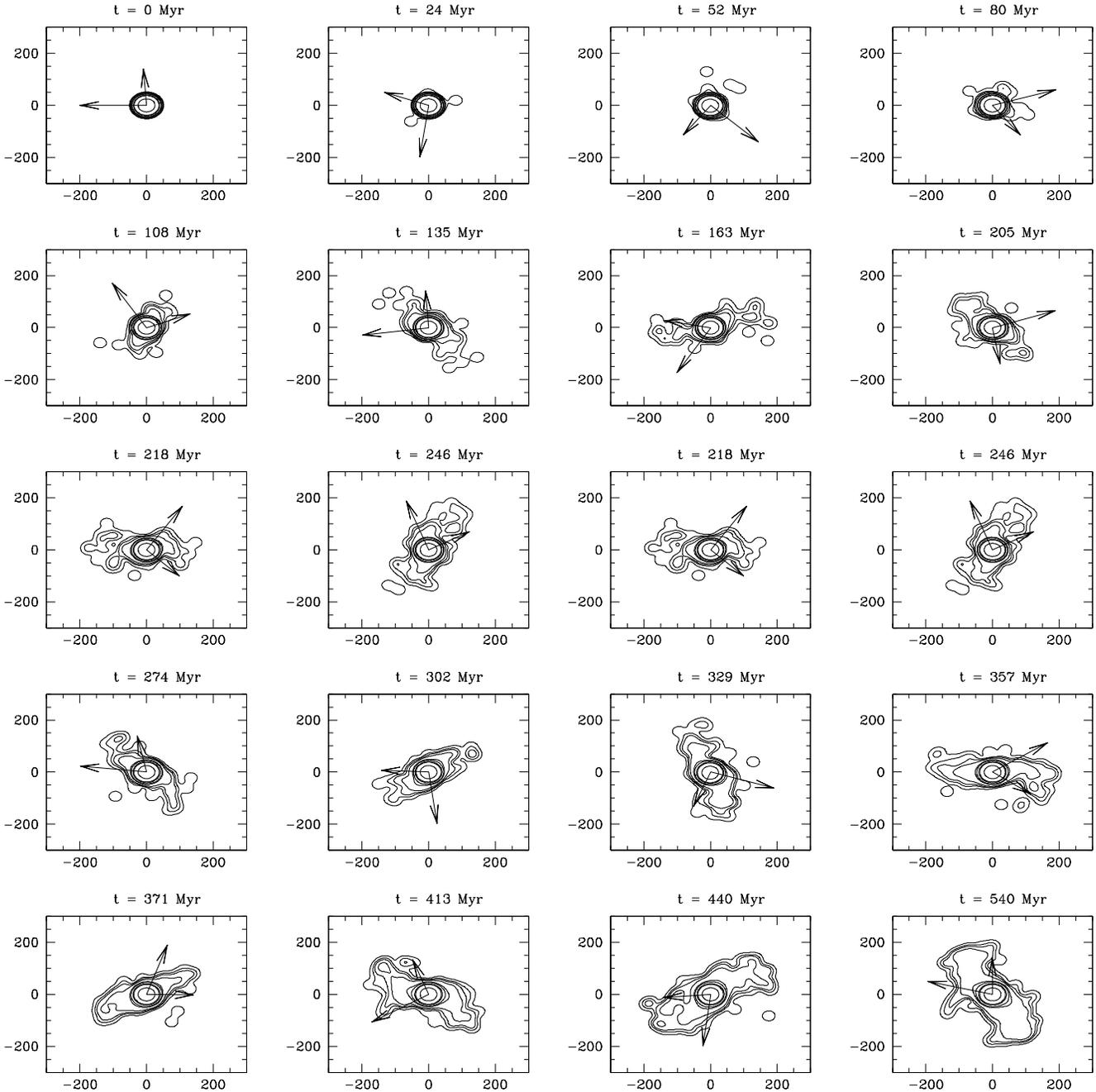}
 \caption{Development of tidal tails of model m5.
Each time step is marked as Myr. Two arrows of each panel indicate
the directions of the galactic centre (long) and its orbit
(short). The unit of coordinate is pc.}
\end{figure*}

\section{Results}

\subsection{Mass Segregation}

During the dynamical evolution of the globular cluster, the stars
in the cluster try to achieve energy equipartition, so that all
stars have the same kinetic energy. It causes the process that
higher mass stars slow down and sink towards the central region of
the cluster, whereas the lower mass stars have higher velocities
than the mean and tend to occupy the outer region of the cluster.
This process is called mass segregation (Spitzer 1987). Many
observational studies have shown the mass segregation effect in
many globular clusters. As a result of mass segregation, the
envelope of a cluster is preferentially populated by low-mass
stars, so they escape from the cluster more easily than the
high-mass ones. Unlike our work, Combes et al. (1999) have
employed models with low mass stars being more populated in the
outer parts to mimic the clusters that have reached the mass
segregation, since their models do not include the two-body
relaxation effects. In the present study, we used models which
have no initial mass segregation but evolve towards the segregated
state automatically. We can see the process of mass segregation in
the variation of mass functions (MFs) of globular clusters with
time. The MF can be represented by a power law shown in the
previous section. The slope of mass functions becomes steeper in
the outer part of the cluster as a result of mass segregation.

Fig.~2 shows the MFs of model m3 at a few selected epochs. In the
initial models, MFs are the same in all regions, but the evolved
models clearly show that the MFs depend on the position in the
cluster. We can see that the massive stars are more concentrated
to the inner region while the less massive stars are dominant in
the outer region. As a result, the MFs of outer region are steeper
than those of inner region. We can also see that the mass
segregation effect increases with time. Because of preferential
migration of high-mass stars towards the centre, the slope of MF
from the inner region tends to be more flattened with time.
However the slope of MF from the whole region does not change very
much until $M \sim 0.5M_i$, where $M_i$ is initial mass of a
cluster (Lee, Fahlman \& Richer 1991; Takahashi \& Lee 2000).
Since the main purpose of our simulation is to examine the
development and evolution of tidal tails, we did not continue the
simulations towards the significant loss the total mass. Thus the
MF of the entire cluster does not change significantly until the
end of our simulations, but the outer parts of the clusters become
quickly dominated by low mass stars since the mass segregation
takes place much faster than the evaporation time-scale.

\subsection{Radial Density Profile}

Indications of tidal tails around globular clusters have been
found in several previous studies of globular cluster radial
density profiles. For a given tidal field, the outer part of a
globular cluster has more stars than that of a King model. As a
result, the slope of the radial density profile shows a deviation
from the King model around the tidal radius of the cluster. We
show some of the examples in Fig.~3. The initial profiles are
shown for comparison.

The radial distribution of extra-tidal stars can be described by a
power law $\Sigma (r) \propto r^{\gamma}$. We derived the average
value of $\gamma = -3.2$ from the simulations. This value is in
well accordance with Combes et al. (1999) who deduced a slope
around $\gamma = -3$ from their N-body simulations. The reason for
such a steep slope is that the stars in the tidal tails are
preferentially on radial orbits, and the gravitational potential
is almost Keplerian. It has long been known that the density
profile falls like $r^{-3.5}$ in such a case.

On the other hand, the observed globular clusters tend to have
shallower slopes. The observed studies of Grillmair et al. (1995)
and Leon et al. (2000) showed that most of the slopes are around
$-1$. Lee et al. (2003) found that the slope of Galactic globular
cluster M92 is $\gamma = -1.27$ from the deep CCD photometry.
Combes et al. (1999) explained the shallower slope in the
observations as the contamination of background-foreground
populations. Although this must be a main source of discrepancy,
we could consider other reasons. As an example, Lee et al. (2003)
showed that the slopes of the density profiles for stars in
different magnitude bins have different values in the outer part.

Testa et al. (2000) obtained the surface density profile of M92
using plates from the Digitized Second Palomar Sky Survey. They
fitted the extra-tidal profile of the cluster to a power law and
found $\gamma = -0.85\pm0.08$. Lee et al. (2003) derived $\gamma =
-0.82\pm0.10$ using only the bright stars ($18.5 < V < 20.5$),
which correspond with the stars in the study by Testa et al.
(2000) from surface density profile of M92. Whereas, using the
faint stars ($22 < V < 23.5$), they found a steeper slope of
$\gamma = -1.31\pm0.09$, which is significantly different from
that of the bright stars.

We found similar trends in our results of N-body simulations. In
Fig.~4, we show an example radial density profiles for different
masses in model m9. We can see that the profile of massive stars
have shallower slope than that of less massive stars. The profile
of massive stars ($2M_\odot < M < 10M_\odot$) follows a slope of
$\gamma = -2.4$ while that of less massive stars ($1M_\odot < M <
2M_\odot$) have a much steeper slope of $\gamma = -3.1$. This
implies that the difference in slopes of the density profiles in
different magnitude bins is a real feature of the clusters. Most
of observational studies have used only bright stars, because of
the sensitivity of the photometry. This could be a part of the
reason for the shallower density distribution in observational
data. The velocity anisotropy depends on the mass of the stars,
and the higher mass stars tends to be on more tangential orbits
than lower mass stars (see, for example, Takahashi \& Lee 2000).
The stars with less radial anisotropy would be populated in
shallower profile.

Since the tidal tails of globular clusters are preferentially
formed by the lowest mass stars (Combes et al. 1999), it is
difficult to study tidal tails of globular clusters using only
bright stars. This means that the study of globular clusters tidal
tail should be performed using stars with the lowest possible mass
that is detectable. Deep CCD studies of other nearby clusters are
needed to verify whether the mass dependence of the density
profile slope is a common phenomenon.

\begin{figure}
\includegraphics [width=84mm] {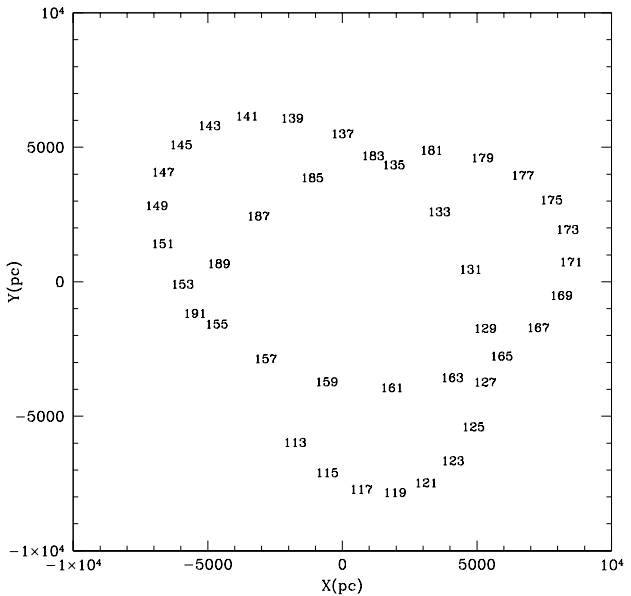}
 \caption{The orbit and location of each step of model m3.
Step numbers are marked.}
\end{figure}

\begin{figure*}
\includegraphics [width=180mm] {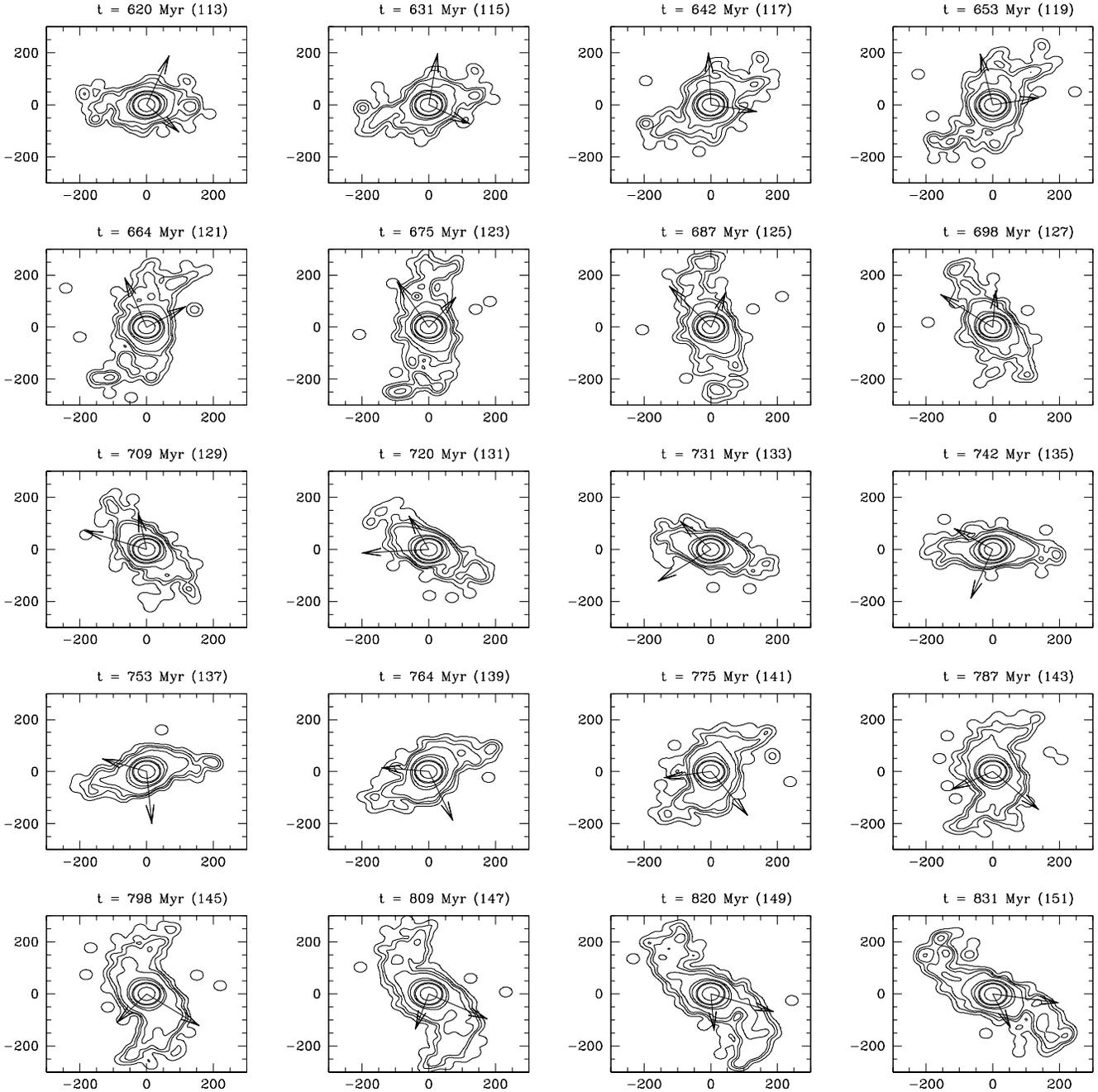}
 \caption{Contours of levels of model m3. Each time-step is marked
as Myr and step number. The positions of each time step in the
orbit can be traced by step numbers in Fig.~6. Two arrows of each
panel indicate the direction of the galactic centre (long) and its
orbit (short). Note the change of length of tidal tails. The unit
of coordinate is pc.}
\end{figure*}

\begin{figure*}
\includegraphics [width=180mm] {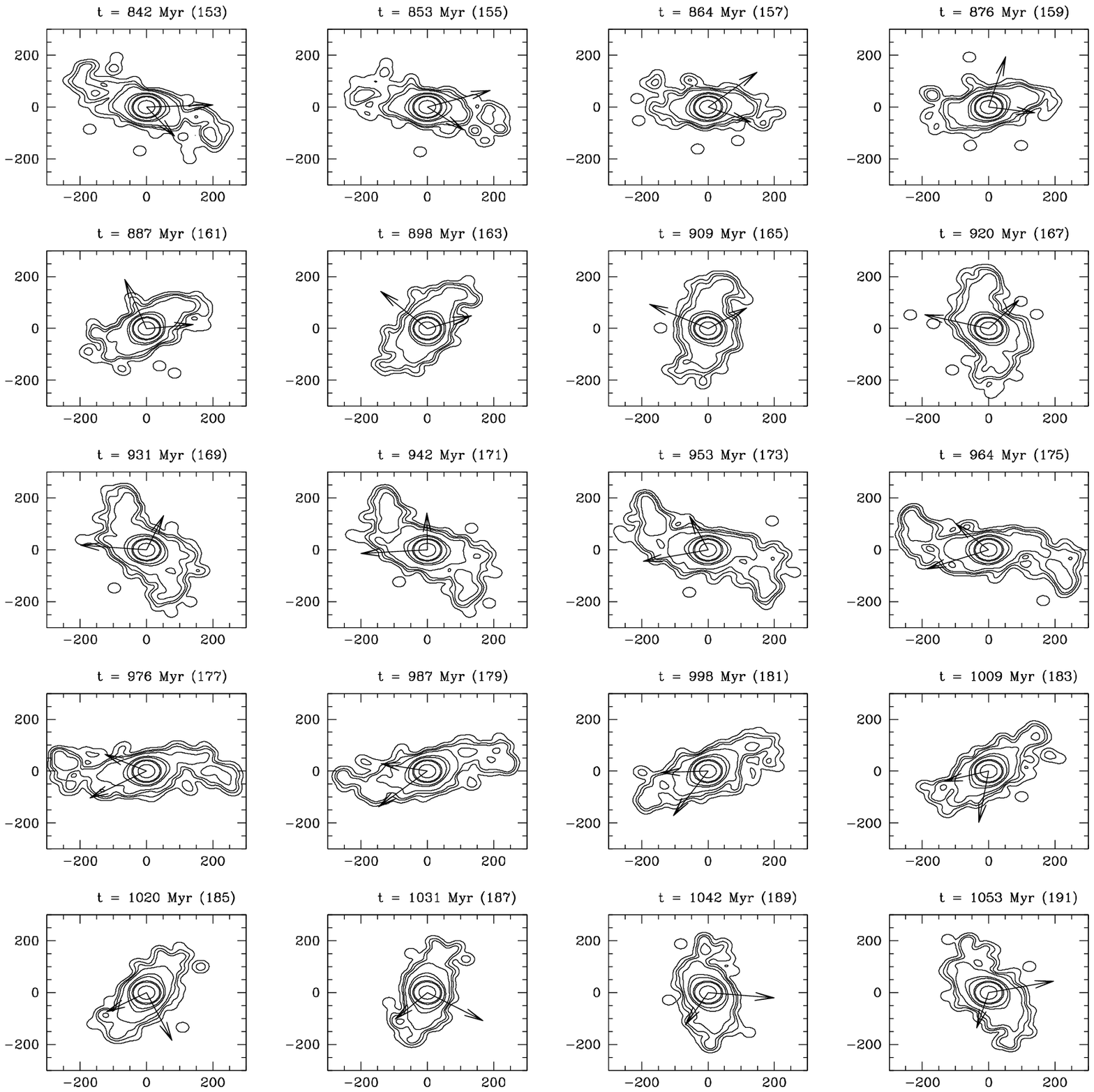}
 \contcaption{}
\end{figure*}

\subsection{Tidal Tails}

The stars unbound from the host cluster do not escape
instantaneously, but they slowly drift from the globular cluster
and form tidal tails. They can form well-defined features outside
the cluster, as observed on the sky (Lee et al. 2003 and
references therein). The development of tidal tails of one model
(m5) is displayed in Fig.~5. Two arrows indicate the direction of
galactic centre (long) and cluster orbital direction (short). The
maps are projected in the (X,Y) plane on which the cluster orbits.
We can see the development of almost symmetrical tidal tails in
two directions.

The directions of tidal tails are known as a good tracer of
cluster orbital phase. Combes et al. (1999) presented from N-body
simulations that there exist two tidal tails around the globular
cluster along its orbit. In large scale, it seems obvious that the
tidal tails would be aligned with the cluster's orbit. Recently,
Odenkirchen et al. (2001, 2003) detected clear tidal tails around
the Galactic globular cluster Palomar 5, extending over an arc of
$10^\circ$ on the sky, corresponding to a projected length of 4
kpc at the distance of the cluster. They also showed that the
tidal streams are aligned with the orbit of this cluster. However,
not all globular clusters show such clear tidal tails like Palomar
5. Actually, Palomar 5 is the first, and only globular cluster
whose tidal tails have been detected at a very significant level
of confidence so far. Most of observational studies about globular
clusters do not cover such a large scale either. In present study,
we focused on a variation of tidal tails of globular clusters
along its orbits in a relatively small scale (within 5$r_t$). We
found that in small scale, the directions of tidal tails of
globular clusters are not consistent with its orbital direction.
In Fig.~6 and Fig.~7, we show the positions in the orbit of one
model (m3) and its contours of levels. To see the direction of
tidal tails, we show the $\theta/\theta_0$ as a function of time
for model m3 in Fig.~8. Here, $\theta_0$ is the angle between the
directions of galactic centre and orbital motion, and $\theta$ is
the angle between the directions of galactic centre and tidal
tail. We can see that the direction of tidal tails is almost
median of the directions of galactic centre and cluster orbit in
all galactic positions.

From these figures, we also find the relation between the length
of the tidal tails and the orbital phase of the cluster. The
cluster which has a circular orbit shows almost the same length of
tidal tails since the strength of tidal tails does not change with
time. However in the elliptical orbit, the lengths of tidal tails
change with time depending on the galactocetric distances $R_G$.
The clusters on elongated orbits experience variation in tidal
field. If the Galactic potential behaves like Keplerian, the tidal
force is proportional to $R_G^{-3}$ while the logarithmic
potential gives rise to the $R_G^{-2}$ dependence. In any case,
the amount of the variation in tidal force for clusters on
elongated orbits would be significant. We may simply assume that
the tidal tails would be longer when the tidal force is stronger.

However, from Fig.~6 and Fig.~7, we can see that the length of the
tidal tails increases towards the apocentre where the tidal force
from the bulge is weakest (near orbit number 147 and 173 in
Fig.~6) and decreases towards the pericentre. Yim \& Lee (2002)
first noted such behavior, and explained that the stars usually do
not respond instantaneously to the external force. Since the stars
near the tidal boundary have orbital periods similar to the
cluster's orbital time around the galaxy, the effect of the
strongest tidal force appears when the tidal force is weakest.

\begin{figure}
\includegraphics [width=84mm]  {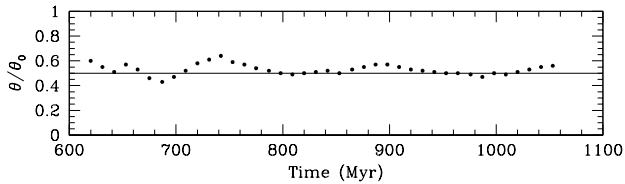}
\caption{Direction of tidal tails of model m3. $\theta_0$ is the
angle between the directions of galactic centre and orbital
motion, and $\theta$ is the angle between the directions of
galactic centre and tidal tail. The direction of tidal tail is
almost median of the direction of galactic centre and cluster
orbital direction. }
\end{figure}

\subsection{Extra-Tidal Stars}

The stars in the tidal tails can be called as extra-tidal stars.
We show the fraction of these stars as a function of time for all
runs in Fig.~9 and Fig.~10. The definition of extra-tidal stars is
clear only when the globular cluster is embedded in a steady
potential, which is the case for a circular orbit. A common
approximation is to assume that a star is unbound when it exceeds
the conventional tidal radius (Meylan \& Heggie 1997), and N-body
models show that this leads to consistent results (Giersz \&
Heggie 1997). However, for the elongated orbits, the tidal radius
varies along the orbit. Keenan (1981) found that the limiting
radii of observed globular clusters are close to the tidal radius
at pericentre. In this study, simply the stars which lie outside
the tidal radius are regarded as extra-tidal stars. For the
elongated orbits, the tidal radius at pericentre have been used as
limiting radius. In calculation of the fraction, the stars
eliminated from the simulations which went beyond the 5 times of
tidal radius were also included in extra-tidal stars.

Generally the fraction increases with time, but the clusters
having elongated orbits show clear decrease at some points (in
models m3, m4, m6, m7, m9 and m12). In Fig.~10, we show the
variation of the fraction of extra-tidal stars of model m9 which
has eccentricity of 0.34. The arrows in Fig.~10 denote the
pericentre in the orbit of m9. We can see that the fraction of
extra-tidal stars reaches small peaks at the pericentre and then
slightly decreases. These points correspond to the time when the
length of tidal tails decreases. We can see that when the length
of tidal tails decreases part of the stars included in tidal tails
are recaptured by the host cluster.

In addition, from the comparison of m2 and m5 in Fig.~9, we can
see the effect of strengths of external potential on the rate of
evaporation of stars. With the same initial cluster model, the one
which has closer orbit from the galactic centre (m5) shows steeper
increase in the fraction of extra-tidal stars. Obviously, the
clusters in stronger tidal fields experience more rapid
evaporation of stars.

\begin{figure}
\includegraphics [width=84mm] {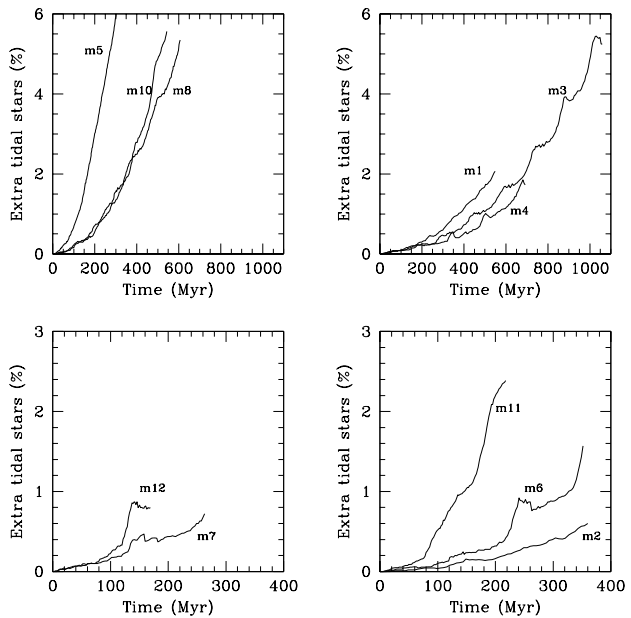}
 \caption{Fraction of extra-tidal stars as a function of time
for all runs except m9. Note that the models having elliptical
orbits show clear decrease at some points. }
\end{figure}

\begin{figure}
\includegraphics [width=84mm] {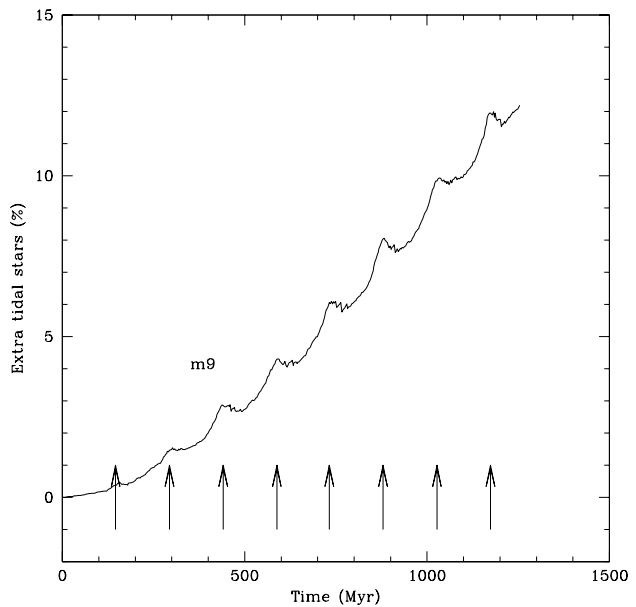}
 \caption{Fraction of extra-tidal stars as a function of time
for model m9. The arrows denote the pericentre in the orbit. The
fraction makes small peaks at the pericentre and then slightly
decreases. }
\end{figure}

\begin{figure}
\includegraphics [width=84mm] {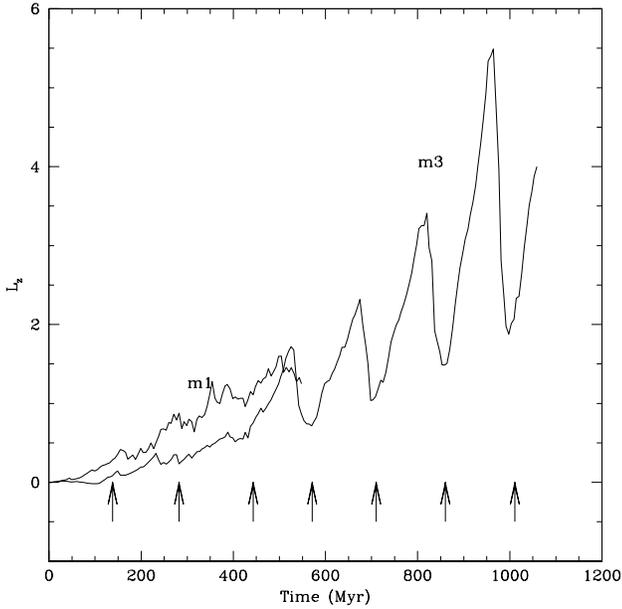}
 \caption{$Z$ component of specific angular momentum of
clusters as a function of time for models m1 and m3. The arrows
denote the pericentre of model m3 having elliptical orbit.}
\end{figure}

\begin{figure}
\includegraphics [width=84mm] {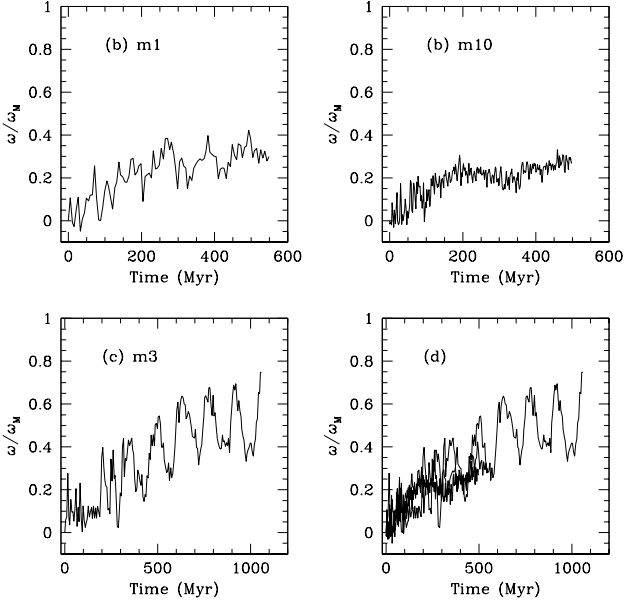}
 \caption{The angular speed of model clusters.
The values are normalized by the maximum angular speed (orbital
angular speed) of each orbit. 3 models are plotted altogether in
panel (d).}
\end{figure}

\begin{figure}
\includegraphics [width=84mm] {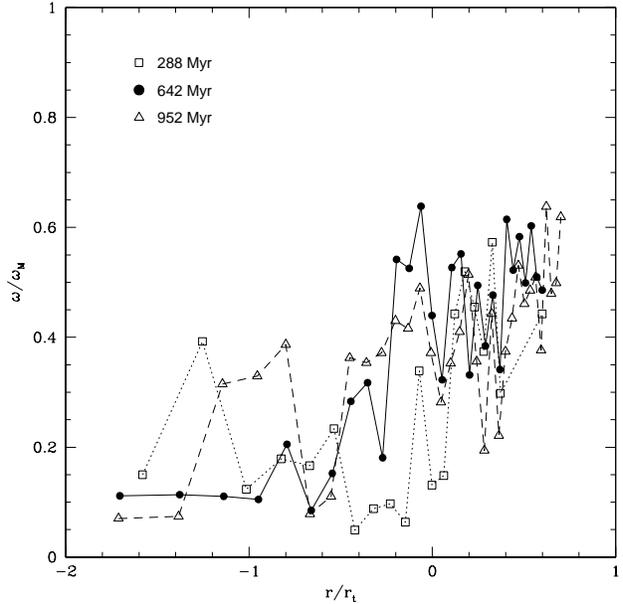}
 \caption{The angular speed of model cluster m3 along the
equator at a few selected epochs. The values are normalized by the
maximum angular speed (orbital angular speed).}
\end{figure}

\subsection{Angular Momentum}

The model clusters used in this study have no angular momentum at
the beginning. However, the probability that the globular clusters
is formed without initial angular momentum is very small. A seed
angular momentum in a proto-cloud would remain in the globular
cluster without dissipation, since the time-scale of cluster
formation is much shorter compared to globular relaxation times.
White \& Shawl (1987) have measured the projected axial ratio
($b/a$) of 100 globular clusters in the Milky Way, where $a$ and
$b$ denote semi-major and semi-minor radii, respectively. They
obtained that mean axial ratio $<b/a> = 0.93 \pm 0.01$. They
argued that the flattened shape of the clusters can be caused by
either anisotropy in velocity dispersion or rotation. Since the
velocity distribution of stars in tidally truncated clusters tends
to become quickly isotropic (Takahashi \& Lee 2000), the
flattening is likely to be caused by rotation (Combes 1999).
Kinematical data also showed that the flattening could be
explained by rotation, and that the minor axes are nearly
coincident with the rotation axes (Meylan \& Mayor 1986).

Einsel \& Spurzem (1999) demonstrated that the influence of
rotation on star cluster evolution is not small. The core-collapse
time could be accelerated significantly by initial rotation.
Though rotation is a natural initial condition from collapse of a
star-forming cloud, it is not easy to include the rotation in most
of the existing evolutionary models of star clusters. Kim et al.
(2002) showed that the evolution of the tidally limited cluster is
significantly accelerated by initial rotation for equal mass
models. Recently, from the multi-mass Fokker-Planck models, Kim,
Lee \& Spurzem (2004) confirmed that the rotation accelerates
core-collapse and dissolution of globular clusters even for
multi-mass models.

The initial angular momentum is expected to disappear from the
cluster with time by the escape of stars carrying some angular
momentum. Frenk \& Fall (1982) showed that the young globular
clusters are flatter on average than old ones in the LMC. Globular
clusters with shorter relaxation time are also rounder (Davoust \&
Prugnel 1990), which supports the loss of rotation with
relaxation. Kim et al. (2002, 2004) showed that the rotational
energy in unit of total energy decreases monotonically with time
in the tidally limited rotating clusters. However, the clusters
could acquire angular momentum through tidal interactions with the
Galaxy. We show the $Z$ component of specific angular momentum
(angular momentum per unit mass) of clusters as a function of time
for models m1 and m3 in Fig.~11. The arrows in Fig.~11 denote the
pericentre of model m3 having elliptical orbit. The two models
have the same initial conditions except for their orbits. Fig.~11
shows that the clusters gradually acquire angular momentum by
tidal torque. The model m1 having circular orbit (maintaining
closer distance from the Galactic centre) has steeper increase in
angular momentum. In model m3 having elliptical orbit, just after
the perigalactic passages, the cluster angular momentum shows
rapid increase. From these results, we can conclude that the
Galactic tidal field play an important role in providing angular
momentum to the clusters. The peaks of angular momentum appear
near apocentre where the length of tidal tails is maximum.

The maximum angular speed acquired by tidal torque would be the
synchronous speed, which means the cluster rotates at the same
angular speed as the orbital angular speed. We show the angular
speed normalized by maximum angular speed (orbital angular speed)
at the tidal radius of clusters as a function of time for three
models in Fig.~12. We can see that the angular speed increases
with time, but our simulations do not last long enough to achieve
the synchronous rotation. When the results for three difference
orbits are folded together as shown in Fig. 12(d), the behavior of
angular speed evolution appears nearly the same for all. We can
not find any clear difference between orbits (Fig.~12(d)) possibly
because of short time coverage. In case of the model m3 (covering
the longest time), the angular speed would reach the maximum in
about 7 times orbital period if it linearly increases.

If a fully relaxed globular cluster has a significant rotation it
would be the result of tidal interactions with the Galaxy because
the initial rotation decrease with time and finally disappear (Kim
et al. 2002, 2004). Lee et al. (2004) suggested that the globular
cluster NGC 7492 which has significantly flattened shape has been
much affected by tidal shocks from the relatively flat MF slope
and its small mass. If the flattened shape of NGC 7492 is caused
by its rotation, Galactic tidal field must have given important
influences. When we get the radial velocity distribution, we will
be able to verify the role of tidal field in rotating of this
cluster.

From the radial profiles of the rotational velocity Kim et al.
(2002, 2004) showed that the rotational velocities of initially
rotating clusters decreases beyond the half-mass radius. As an
observational result, the rotation of the bright Galactic globular
cluster $\omega$ Centauri is almost solid-body until about 15\% of
the tidal radius, and then falls off quickly (Meylan \& Mayor
1986, Merritt, Meylan \& Mayor 1997). However, we found that the
rotating clusters accelerated by tidal torque have different shape
of radial angular speed distribution from initially rotating
clusters. In Fig.~13, we have shown the distribution of angular
speed as a function of radius for model m3. Although the data are
rather noisy, the angular speed starts from a differential
rotation with higher angular speed at larger radii to rigid body
rotation. This is due to the fact that the tidal torque is more
effective at the outer parts, but the inner parts gets the angular
speed through the relaxation. The rotational energy for a
uniformly rotating stellar system with angular speed $\omega$ is
around $M r_t \omega^2$ and maximum value of $\omega$ obtained by
the tidal torque is $\sim v_{rot}/R_G$, where $v_{rot}$ is the
velocity of galactic rotation. The ratio of rotational kinetic
energy to gravitational potential energy can thus become
$T_{rot}/W\sim \left({r_t\over R_G}\right)^2 \left( {v_{rot}\over
\sigma}\right)^2$, where $\sigma$ is the velocity dispersion of
the cluster. For a cluster at $R_G=5$ kpc, $r_t=25$ pc, and
$\sigma=5$ km/s,  $T_{rot}/W$ could be about 0.05. Such an amount
of rotation could produce the dynamical ellipticity greater than
0.1 (e.g., Kim et al. 2002).

\subsection{Velocity Anisotropy}

\begin{figure}
\includegraphics [width=84mm] {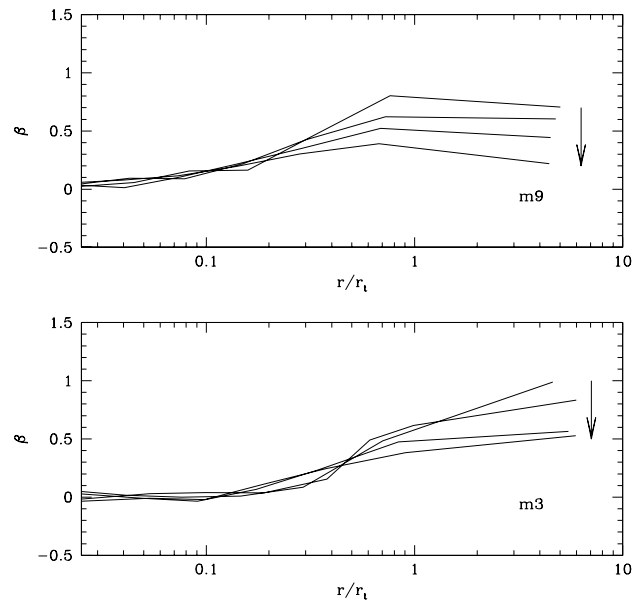}
 \caption{The radial variation of anisotropy parameter of models
m1 and m3. m1 has time range of $132 \sim 554$ Myr, and m3 has
$133 \sim 1065$ Myr. All models show radial anisotropy in the
outer parts of the clusters and the radial anisotropy decreases
with time. }
\end{figure}

It has long been known that isolated star clusters become radially
anisotropic after core-collapse due to the scattering of stars out
of the cluster core (Spitzer 1987). The development of velocity
anisotropy was investigated in some works which used various
numerical methods. Most of these calculations confirmed the
earlier expectations of generation of radial anisotropy in the
outer parts of the clusters, although the detailed behavior
depends on the numerical methods. However, actual clusters are
embedded in the Galactic tidal field which should impose a finite
boundary to the cluster. The introduction of tidal boundaries
gives significant effects on the velocity anisotropy because the
stars could outflow over tidal boundary. Giersz \& Heggie (1997)
found in N-body simulations that star clusters in tidal fields
remain isotropic during their evolution except for the outer parts
which become tangentially anisotropic due to the preferential loss
of stars on radial orbits. Takahashi \& Lee (2000) found in their
anisotropic Fokker-Planck simulations that the clusters develop
stronger radial anisotropy in the outer part of the cluster during
the early phase and then the development of  radial anisotropy
stops, and finally strong tangential anisotropy appears as the
cluster loses mass. Recently, Baumgardt \& Makino (2003) showed
with N-body simulations that the outer parts of the clusters
become rapidly tangentially anisotropic.

Though our simulations do not cover enough time of cluster mass
loss, it might be interesting to look for signs of velocity
anisotropies in our runs. Following Binney \& Tremaine (1986), we
define an anisotropy parameter $\beta$ as,
\[
\beta = 1 - {{v_t^2} \over {2v_r^2}},
\]
where $v_r$ and $v_t$ are the radial and tangential velocity
dispersions respectively. They are measured in a non-rotating
coordinate system in which the cluster centre is always at rest at
the origin.

Fig.~14 shows the radial variation of $\beta$ for models m9 and m3
at various times. All models show radial anisotropy in the outer
parts of the clusters. The velocity anisotropy is expected to be
changed with time. We can see that the radial anisotropy decreases
with time. Extra-tidal stars also show radial anisotropy and do
not show clear difference from outermost parts of bounded stars.
From Fig.~14, we also confirm the development of radial anisotropy
in the outer parts, and that the radial anisotropy decreases with
time. Except for the outermost parts, clusters are nearly
isotropic and there is no sign of the radially anisotropic
velocity dispersion which forms in isolated clusters. Mass
segregation could be a reason. Since massive stars are more
centrally concentrated, they tend to be located at lower energy
levels which have more isotropic velocity distributions (Baumgardt
\& Makino 2003).

\section{Summary and Conclusions}

We have carried out a dozen N-body simulations of the globular
clusters within the external tidal field. Our main conclusion can
be summarized as follows:

\begin{enumerate}
\item All runs show very quick development of tidal tails within a
few orbits.

\item Mass segregation effect appears in all runs while the
initial mass function is still being preserved. Since the rate of
stellar evaporation depends on the mass of the stars, the mass
function would change substantially when the amount of evaporated
stars becomes more than 50\% of the initial mass.

\item The radial distribution of extra-tidal stars can be
described by a power law with a slope around $-3.2$. This result
is well consistent with previous study (Combes et al. 1999). In
addition, the radial profile of massive stars have shallower slope
in the outer part of the cluster than that of less massive stars
as observed in M92 (Lee et al. 2003). This could be a part of the
reason why the observational data have shallower slopes than
predicted by simulations.

\item The directions of tidal tails are determined by the location
of the cluster and its orbits. The length of tidal tails increases
towards the apocentre and then decreases towards pericentre. When
the length of tidal tails decreases some of the stars in the tails
are recaptured by the host cluster. In addition, we verified that
the external potential plays an important role in destruction of a
cluster.

\item The clusters which have no initial angular momentum gain
angular momentum from tidal interactions with galaxy. Angular
momentum of model clusters gradually increases with time. The
model clusters having elongated orbits show rapid increases in
angular momentum just after the perigalactic passages.

\item The rotating clusters accelerated by tidal torque start with
the rotational angular speed decreasing towards the centre but
evolve towards the rigid-body rotation.

\item In the early stages of globular cluster evolution, the
clusters have radial anisotropy in the outermost parts, while
clusters are nearly isotropic in the central region. And the
radial anisotropy decreases with time.

\end{enumerate}

These N-body simulations help to understand the recent
observations of extended tidal tails around globular clusters. In
this study we did not consider tidal shock by galactic disk for
simplicity. However disk shock also plays important role in the
evolution of globular clusters (Gnedin et al. 1999). Disk
component should be contained in subsequent studies to reproduce
more realistic Galaxy models. Although we used many particles to
reproduce real clusters as possible, it is still far from the real
ones. It is necessary to perform N-body simulations with large
number of stars up to $N \sim 10^5$.

\section*{Acknowledgments}

This work was supported by KRF grant No. 2002-041-C00123.
Numerical calculation was carried out by supercomputers at KISTI
trough Grand Challenge Program. H.S. acknowledges the support of
the Korea Science and Engineering Foundation (KOSEF) to the
Astrophysical Research Center for the Structure and Evolution of
the Cosmos (ARCSEC$''$) at Sejong University.

\label{lastpage}

\end{document}